\RequirePackage[hyphens]{url}

\documentclass[
]{ceurart}

\usepackage{graphics} 

\usepackage{listings}
\begin{document}

\copyrightyear{2024}
\copyrightclause{Copyright for this paper by its authors.
  Use permitted under Creative Commons License Attribution 4.0
  International (CC BY 4.0).}

\conference{INRA'24: 12th International Workshop on News Recommendation and Analytics in Conjunction with ACM RecSys 2024, 18 October 2024, Bari, Italy}

\title{Enhancing Prediction Models with Reinforcement Learning}


\author[1,2]{Karol Radziszewski}[%
email=karol.radziszewski@ringieraxelspringer.pl,
]
\cormark[1]
\fnmark[1]

\author[1]{Piotr Ociepka}[%
email=piotr.ociepka@ringieraxelspringer.pl,
]
\cormark[1]
\fnmark[1]

\address[1]{Ringier Axel Springer Polska, Warsaw/Krak\'{o}w, Poland}
\address[2]{Warsaw University of Technology, Warsaw, Poland}

\cortext[1]{Corresponding author.}
\fntext[1]{These authors contributed equally.}

\begin{abstract}
  We present a large-scale news recommendation system implemented at Ringier Axel Springer Polska, focusing on enhancing prediction models with reinforcement learning techniques. The system, named \textit{Aureus}, integrates a variety of algorithms, including multi-armed bandit methods and deep learning models based on large language models (LLMs). We detail the architecture and implementation of \textit{Aureus}, emphasizing the significant improvements in online metrics achieved by combining ranking prediction models with reinforcement learning. The paper further explores the impact of different models mixing on key business performance indicators. Our approach effectively balances the need for personalized recommendations with the ability to adapt to rapidly changing news content, addressing common challenges such as the cold start problem and content freshness. The results of online evaluation demonstrate the effectiveness of the proposed system in a real-world production environment.
\end{abstract}


\maketitle

\section{Introduction \label{sec:intro}}
Ringier Axel Springer Polska is among the largest media companies in Poland, operating the news website \url{www.onet.pl}. Onet.pl attracts approximately 6 million unique users monthly, representing about 20\% of the Polish internet users \cite{wirtualnemedia}. According to SimilarWeb, Onet is the 18th largest news website globally \cite{similarweb}. Our recommendation system, called \textit{Aureus}, processes over a thousand requests per second, necessitating low latency to ensure users experience minimal wait times for website loading. 

\textit{Aureus} comprises a variety of recommendation components, including user segmentation and reinforcement learning (popularity-based component). Over the past year, we implemented additional modules responsible for content similarity and deep learning models based on on large language models (\textit{LLM}) to capture individual preferences.

In this article, we focus on describing the architecture of a real-world large-scale news recommendation system, in particular:
\begin{itemize}
\item we show that combining ranking prediction models with reinforcement learning significantly improves online metrics,
\item we further analyze different aspects of model configuration and training objectives concerning multiple business KPIs.
\end{itemize}

\section{Related Work \label{sec:related_work}}
The complexity of news recommendation systems exceeds that of other systems due to the rapid data updates and content obsolescence, with thousands of articles published daily. A key challenge is addressing cold start users, as many visitors rely on cookie IDs without logging in.

Traditional collaborative filtering methods, such as matrix factorization \cite{collaborative_filtering}, face difficulties due to the cold start problem, requiring multiple observations for each user. To overcome this, models using external features have been proposed. Wang et al. introduced RippleNet, a deep learning model that leverages external data via knowledge graphs, enabling recommendations with minimal previous users' interactions \cite{ripplenet}.

Reinforcement learning, particularly multi-armed bandit algorithms, is another solution to cold start. This approach has been successfully used in our systems for several years \cite{inra_rasp19}. 

The subsequent methodology of modelling recommendation systems, content-based filtering is crucial, particularly in news recommendations. Recent Natural Language Processing (NLP) advances, such as pretrained models like GPT \cite{openai_embeddings} and PolBERT \cite{Kleczek2020}, have enhanced the generation of personalized recommendations through embeddings.

However, these models are often large and costly. Wu et al. addressed this by introducing NewsBert \cite{wu2021newsbertdistillingpretrainedlanguage}, a distilled version of BERT tailored for the news domain, reducing model size and complexity.

The core aim of recommendation systems is to boost user satisfaction, often measured by clicks or time spent on the platform. While click modeling is straightforward, time-based metrics are more complex. Covington et al. proposed a method to weight clicks based on time spent, implemented in YouTube’s system \cite{youtube_recsim}.

Our approach uniquely integrates bandit algorithms with traditional ranking models, creating an adaptive news recommendation engine that combines the strengths of both multi-armed bandits and deep learning models within a unified architecture.

\section{Proposed Approach}

Over time, \textit{Aureus} has expanded to incorporate a variety of recommendation algorithms and methods, each characterized by distinct capabilities and limitations. This section provides a detailed overview of several of these methods, followed by the introduction of a novel approach for aggregating multiple recommendations into a unified output. This approach leverages the unique strengths of each constituent algorithm while effectively mitigating the specific drawbacks associated with individual methods.

\subsection{Reinforcement Learning}
The initial application of \textit{Aureus} was to automate the curation process for the \textit{Onet.pl} news feed. The method required the capability to rapidly collect user feedback, identify both short- and long-term popularity trends, and recommend content that was both highly popular and engaging. Additionally, the system needed to adapt to emerging articles as well as those experiencing a decline in user engagement over time. Given the primary objective of automating the existing editorial workflow, the recommendations were designed to be population-wide, independent of individual user preferences or tastes.

\subsubsection{Multi-armed Bandits}
Considering the outlined requirements, we selected multi-armed bandit algorithms as the foundation of our approach. This class of methods is particularly suited for balancing the trade-off between exploration -- acquiring knowledge regarding each article's performance and popularity -- and exploitation -- recommending the highest-performing content. Moreover, multi-armed bandit algorithms possess the capability to optimize a wide range of business-related Key Performance Indicators (KPIs), including both continuous and discrete metrics. This flexibility makes them an ideal choice for the dynamic and demanding environment of the publishing industry. Following extensive offline and online evaluations, we identified \textit{Upper Confidence Bound} \cite{ucb} and \textit{Thompson Sampling} \cite{thompson} as the most effective bandit methods for this application.

Nevertheless, the exclusion of individual user preferences emerged as a significant limitation of the selected approach. To overcome this constraint, while preserving the robustness, simplicity, trend-responsiveness, and cost- and time-efficiency of the bandit-based recommender system, we introduced the concept of user segmentation.

\subsubsection{Users Segmentation}
Segmentation involves dividing the entire user population into smaller, more homogeneous groups, each consisting of users with similar tastes. By applying multi-armed bandit algorithms separately within each segment, the recommendation process remains primarily popularity-based. However, through segmentation, each user is presented with a set of articles that are most popular among individuals with comparable interests, thereby enhancing the overall user experience.

The initial approach to user segmentation was based on topic modeling \cite{inra_rasp19}. Specifically, each article was transformed into a simplified embedding using \textit{Latent Dirichlet Allocation} (LDA). Subsequently, user interest profiles were generated by averaging the LDA embeddings of the articles read by each individual user. Then, user interest profiles were clustered using the \textit{k-Means algorithm}.

Although successful and effective, this method was soon enhanced by substituting LDA modeling with \textit{Item2Vec} embeddings \cite{item2vec}. This modification significantly simplified and accelerated the segmentation process by eliminating the need for text analysis, thereby rendering the method language-agnostic. Consequently, this improvement allows for the deployment of \textit{Aureus} across digital publishers regardless of the language in which they publish.

\begin{figure}[h]
    \centering
    \includegraphics[width=\linewidth]{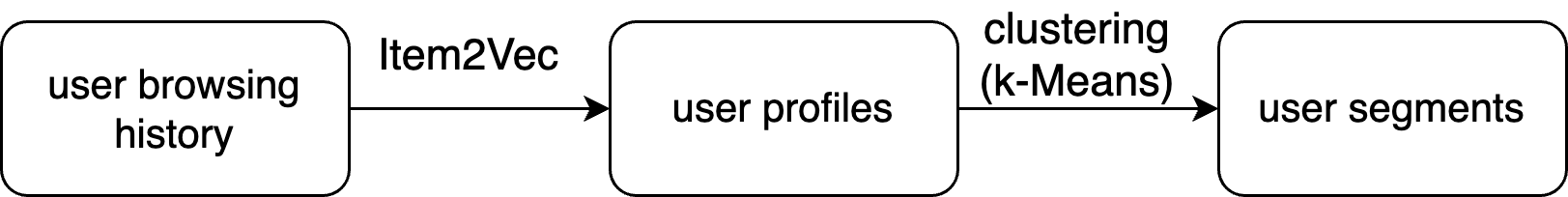}
    \caption{The diagram of the segments calculation process.}
    \label{fig:segmentation_flow}
\end{figure}

\subsection{Prediction Model}
Our models are based on articles that a user has read within the last \textit{N} days. In our experiments, we use an arbitrary value of \textit{N} = 30 days. We calculate user representation by averaging the embeddings of these articles, created by already pretrained LLM PolBert model \cite{Kleczek2020}. Subsequently, we develop two types of models:
\begin{figure}[h]
    \centering
    \includegraphics[width=\linewidth]{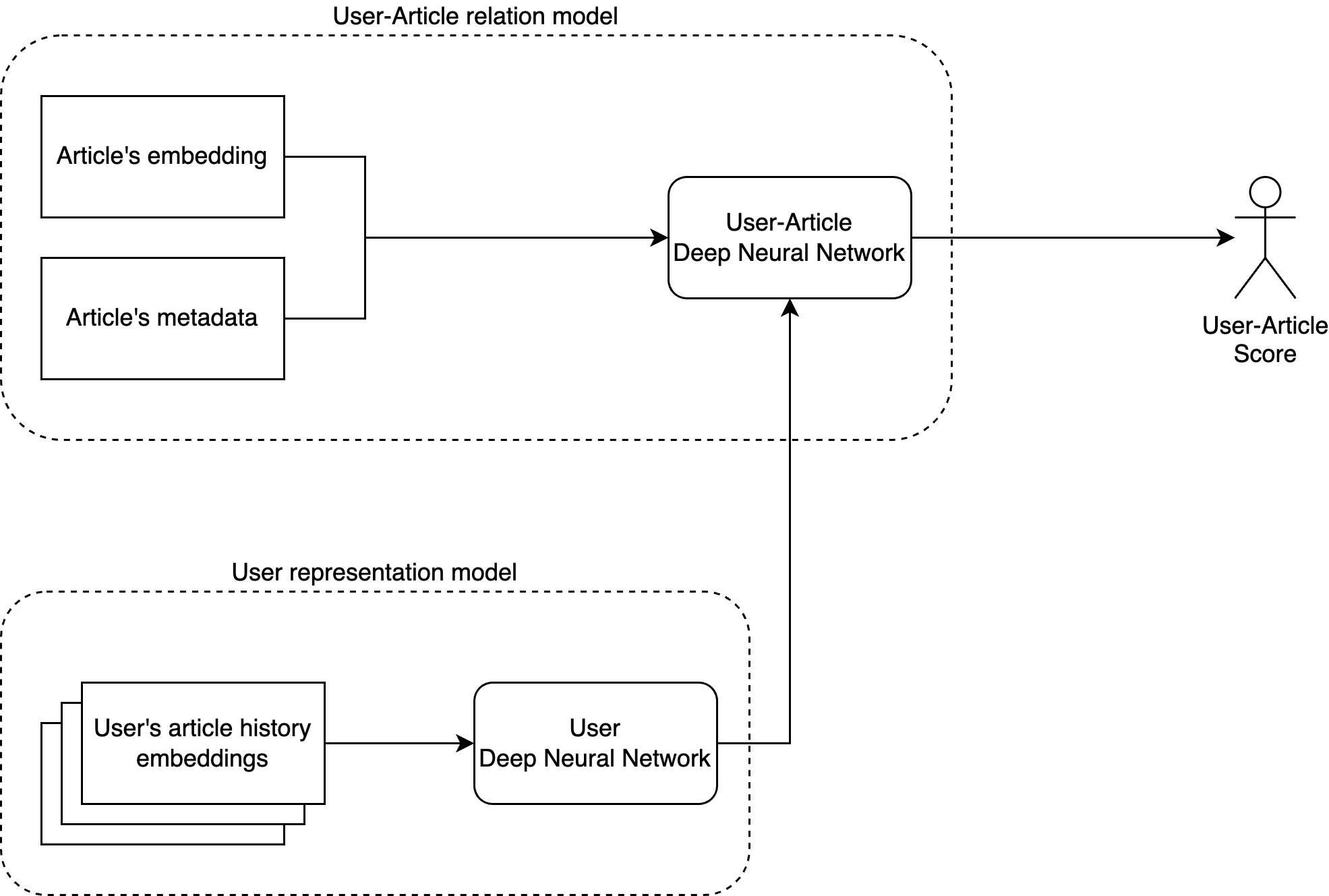}
    \caption{The diagram of the deep model architecture. Input embeddings are calculated with pretrained models. }
    \label{fig:deep-model}
\end{figure}

\subsubsection{Similarity Model}
A simple model that compares user embedding to article embeddings using cosine similarity. This was our initial approach.
\subsubsection{Deep Model}
We created a trainable model that was trained on user clicks as a target variable. Given our large and imbalanced dataset, we sampled an equal number of clicked and unclicked articles to ensure balanced data for evaluation. Using neural network architecture drawn in Figure \ref{fig:deep-model}, we seamlessly integrated additional features into our recommender system, such as article length and other parameters. Since our business KPI is a continuous variable, we also trained models with clicks weighted by this KPI, similar to the approach described in \cite{youtube_recsim}. Weighting by the business KPI resulted in an increase in this KPI in online tests.

\subsection{Model Ensemble Architecture}
We previously outlined two key components of our recommendation system. The reinforcement learning module identifies popular and trending articles, while the prediction model captures individual user preferences. For optimal user satisfaction, the recommendation system must integrate both aspects. Relying solely on a popularity-based model neglects individual user preferences, whereas a user-preference model may overlook trending articles, which are crucial in the news domain.

We evaluated several methods for combining multiple recommendations, two of which advanced to the online testing phase and are now employed in daily operations:

\begin{itemize} 
  \item \textbf{Proportional Random Mixer} — In this approach, each recommendation method is assigned a target proportion within the final content set (e.g., 40\% of recommendations from a bandit algorithm and 60\% from a deep learning model). For the 
\textit{k}-th position in the final recommendation list, an article is selected randomly from the \textit{k}-th positions of the input recommendations, with the selection probability proportional to the assigned target share. 
  \item \textbf{Weighted Average Mixer} — In this method, each content item from the input recommendations is associated with a score from the corresponding model. These scores are normalized to the range \textit{[0.0, 1.0]} to ensure equity. Each content item is then assigned a new score, which is a weighted average of the scores from the input models, and the final recommendation list is ordered based on these new scores.
\end{itemize}

Online testing proved that the weighted average mixer performed significantly better. Consequently, all results and conclusions presented in this paper are based on the weighted average mixer.

\begin{figure}[h]
    \centering
    \includegraphics[width=\linewidth]{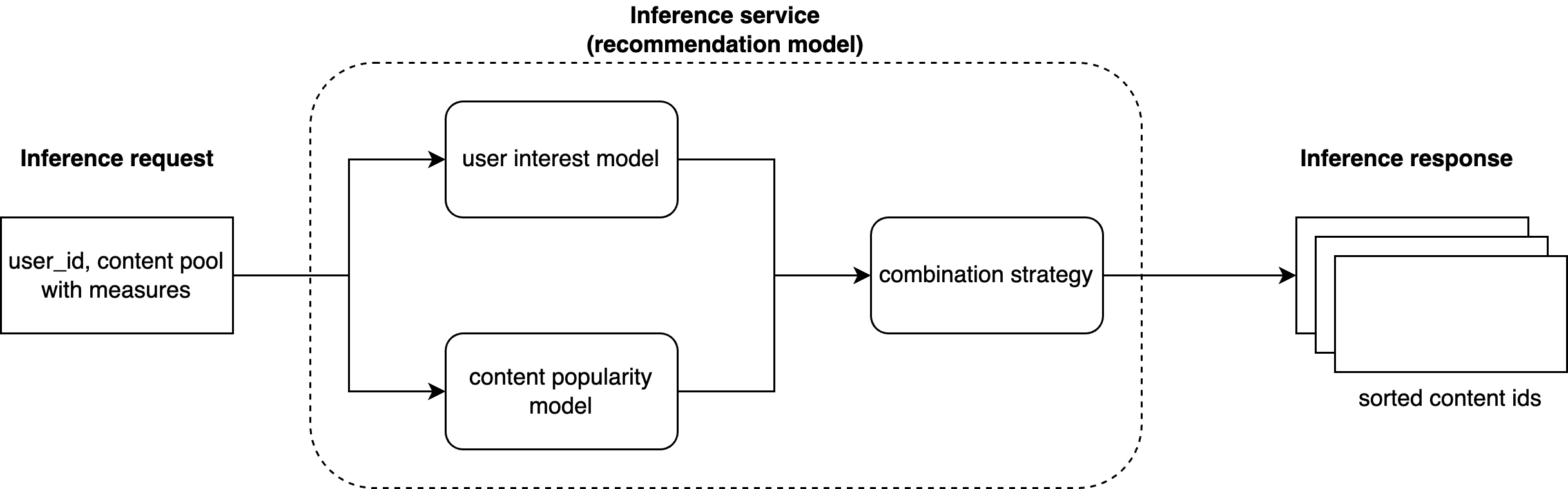}
    \caption{The diagram of the Aureus recommendation system illustrates the following components: Inputs consist of the user ID, a set of content items, and online business KPI metrics. The system integrates two submodels: a deep learning-based user interest model and a multi-armed bandit content popularity model. These submodels are combined using a specified combination strategy.}
    \label{fig:individualizaion_model}
\end{figure}

\section{Experimental Evaluation}
In this section, we present a comprehensive evaluation of our proposed recommendation system. The evaluation is divided into three subsections: Offline Evaluation, Online Evaluation, and Results.

Offline Evaluation describes the performance metrics derived from historical data, allowing us to assess the model’s predictive accuracy in a controlled environment. This process helps identify the best model for subsequent online testing.

Online Evaluation involves deploying the model in a live setting on Onet.pl, where we measure its real-time effectiveness, user engagement metrics and buisness KPI metrics.

Finally, the Results subsection synthesizes the findings from both offline and online evaluations.

\subsection{Offline Setup}
\subsubsection{Baselines}
Each time we develop a new architecture or introduce a new feature, we evaluate the models against at least two baselines:
\begin{itemize}
    \item \textbf{random model} — If our model does not outperform random recommendations, we conclude that it is unsuited for production deployment.
    \item \textbf{current production model} — Our goal is to match or exceed the results of the current production model. If the new model achieves comparable or better results, we proceed to test it in the online environment.
\end{itemize}

\subsubsection{Offline Evaluation Metrics}

We implemented both standard and custom ranking evaluation metrics on historical data. Each metric is calculated with different values of \textit{k} (primarily 3, 5, 10, 15, and 30). Our goal is to train a single model that can be deployed for an extended period. Therefore, we validate our model on three different days: one day, seven days, and thirty days after training. We utilize the following metrics:
\begin{itemize}
    \item \textbf{Standard Metrics} — \textit{NDCG}, \textit{Precision}, \textit{Recall}, \textit{Coverage} and \textit{AUC}
    \item \textbf{Custom Metrics} —  We aim to optimize a continuous buisness KPI with a click prediction model, so we calculate the average value of this KPI for a given ranking at \textit{k} as our custom metric. These are:
    
    \begin{itemize}
        \item \textit{Average Label Value} — This metric considers all articles in the list.
        \item \textit{Average Positive Label Value} — This metric considers all articles that are clicked.
    \end{itemize}
\end{itemize}
\subsection{Online AB Tests}
\subsubsection{Testing Setup}
A critical component of the \textit{Aureus} system, alongside the recommendation models, is the A/B testing engine. This engine facilitates statistically significant and fair online testing of multiple recommendation approaches. Users are randomly and stably assigned to one of the testing variants, independent of user agent, demographic factors, or other variables that might influence the test results. During the test period, each user is exclusively presented with recommendations generated by the model associated with their assigned variant. Key performance indicator (KPI) values for content are collected and recorded according to the testing variants, enabling subsequent analysis and comparison.

It is important to note that the online tests presented in this paper were conducted on a curated sample of users and focused specifically on a designated section of the webpage (recommendations displayed beneath articles). As such, the results may not fully generalize to similar experiments conducted under different conditions or in other areas of the webpage.

\subsubsection{Online Monitoring}
When the model is deployed in a production environment, we continuously monitor its performance with respect to business KPIs and latency. We enforce a stringent latency threshold, beyond which the recommendations generated by the model would not be utilized. To track these online metrics, we employ AWS QuickSight for business-related metrics and Grafana for technical metrics.

\subsection{Results}
Table \ref{table:offline_metrics} presents the offline performance metrics of three predictive models: the random baseline, the similarity model, and the deep learning model. The deep learning model demonstrates superior performance, surpassing the similarity model by approximately 65.7\% in NDCG, around 16.3\% in AvgLabelValue and around 0.9\% in AvgPositiveLabelValue. This indicates that the deep learning model is more effective in terms of business (KPIs) and has been deployed in online tests as the user-to-item recommendation model.

\begin{table}[h]
\caption{Offline metrics for recommendations from the week after the last training day.}
\begin{tabular}{r|c|c|c}
\textbf{model} & \textbf{NDCG@5} & \textbf{\begin{tabular}[c]{@{}l@{}}AvgLabel\\ Value@5\end{tabular}} & \textbf{\begin{tabular}[c]{@{}l@{}}AvgPositive\\Label\\ Value@5\end{tabular}} \\ \hline
random & 0.004 & 0.313 & 60.341 \\ \hline
similarity & 0.035 & 2.793 & 61.551 \\ \hline
deep & 0.058 & 3.248 & 62.112       
\end{tabular}
\label{table:offline_metrics}
\end{table}

We implemented our models in two production environments: the \textit{Onet.pl} homepage and article pages (with recommendations below each article). The determination of the model that achieves the status of ,,king of the hill'' is based on the results of online testing. This approach allows for an evaluation of not only the model's performance but also its alignment with the actual needs of users. In the following section, we compare several models employed by \textit{Aureus}:
\begin{itemize}
    \item random sample from the set of articles,
    \item Thompson Sampling bandit (our golden standard of recommenders),
    \item Thompson Sampling bandit with user segmentation enabled,
    \item items' cosine similarity to the currently read article,
    \item segmented bandit mixed with item-to-item similarity model,
    \item segmented bandit mixed with user-to-item deep model,
    \item segmented bandit mixed with both item-to-item similarity and user-to-item deep model.
\end{itemize}

For comparison, we use two main metrics:
\begin{itemize}
    \item \textit{uplift} -- the percentage difference between the average business KPI value of pieces of content returned by the tested model and that returned by the baseline model,
    \item \textit{latency} -- the median response time, measured in milliseconds, of the tested model; this auxiliary metric serves as a sanity check to ensure that the news website provides users with reasonably responsive performance.
\end{itemize}

\begin{table}[h]
    \caption{Comparison of the performance of individual and combined recommender models on article pages (all uplifts are calculated against ``bandit (TS)'' variant)}
    \begin{tabular}{r|c|c}
         \textbf{model} & \textbf{uplift} & \textbf{latency} \\
         \hline
         random (baseline) & -35.2\% & 11 ms \\
         \hline
         bandit (TS) & --- & 11 ms \\
         \hline
         segmented bandit (TS) & +21.6\% & 12 ms \\
         \hline
         similar (item-to-item) & +26.6\% & 53 ms \\
         \hline
         \begin{tabular}[c]{@{}r@{}}segmented bandit mixed\\with similar (item-to-item)\end{tabular} & +37.7\% & 57 ms\\
         \hline
         \begin{tabular}[c]{@{}r@{}}segmented bandit mixed\\with deep model (user-to-item)\end{tabular} & +35.2\% & 109 ms \\
         \hline
         \begin{tabular}[c]{@{}r@{}}segmented bandit mixed\\with similar and deep model\end{tabular} & +43.9\% & 115 ms \\
    \end{tabular}
    \label{table:online_metrics}
\end{table}

Table \ref{table:online_metrics} presents the results observed during our online testing process. The data clearly demonstrate the synergy effect of the ensembled models, which consistently outperform the individual models. It is also noteworthy that the offline evaluations differ slightly from the online test results, where the combination of similarity-based methods with bandits slightly outperformed the deep model mixed with bandits. From our experience, this discrepancy is common in the context of news and time-sensitive content, where deep models alone may struggle to capture the temporal dynamics.

In terms of latency, deep models substantially increase the response times of the Aureus recommender. However, this increase remains within acceptable limits and does not negatively impact the user experience. Furthermore, incorporating more than two models in the mixing process does not significantly extend response times.

\section{Conclusions and Future Work \label{sec:summary}}
We demonstrated an enhancement of recommendation systems by integrating multiple models into a unified architecture. This hybrid approach facilitates the seamless incorporation of new recommendation scores, enabling the modeling of diverse recommendation aspects. Future work will involve the exploration of additional features, different mixing strategies and various embedding models to further refine the system.

\bibliography{main}

\end{document}